# استفاده از تابع اغتشاش برای آنالیز ارتعاشی تیر با ترک لبه باز

موسی رضائی[1]، سعید لطفان[2]، وحید عرب‌ملکی[3]

[1] دانشیار گروه مهندسی مکانیک، دانشکده فنی مهندسی مکانیک، دانشگاه تبریز  m_rezaee@tabrizu.ac.ir
[2] دانشجوی کارشناسی مهندسی مکانیک، دانشکده فنی مهندسی مکانیک، دانشگاه تبریز  saeedlotfan@ymail.com
[3] کارشناس ارشد مهندسی مکانیک، دانشکده فنی مهندسی مکانیک، دانشگاه تبریز  vahid_maleki@ymail.com

## چکیده

در این مقاله، مدل ارائه شده توسط Shen و Pierre برای بررسی رفتار ارتعاش عرضی تیر دو سر ساده، مورد بازنگری قرار گرفته و با اعمال فرضیات واقع بینانه‌تر ترک به صورت یک اغتشاش پیوسته مدل شده که با استفاده از مکانیک شکست این اغتشاش به صورت یک تابع ارائه شده است. در ادامه فرکانس‌های طبیعی متناظر با مدل، با استفاده از روش گالرکین استخراج شده سپس اثر پارامترهای ترک بر رفتار ارتعاشی تیر ترک‌دار مورد بررسی قرار گرفته است. نتایج بدست آمده نشان می‌دهد که فرکانس‌های طبیعی تیر با افزایش عمق ترک کاهش می‌یابد. در پایان نتایج بدست آمده با نتایج تجربی مقایسه شده‌اند. نتایج نشان می‌دهد که مدل ارائه شده نسبت به مدل‌های قبلی بهبود یافته و رفتار ارتعاشی تیرهای ترک‌دار را با دقت بهتری به ازای پارامترهای مختلف ترک پیش‌بینی می‌نماید.

## واژه‌های کلیدی

رفتار ارتعاشی، تیر ترک‌دار، فرکانس‌های طبیعی، تابع اغتشاش

## مقدمه

سازه‌های دینامیکی که تحت بارهای متناوب قرار می‌گیرند در ماشین آلات صنعتی بخش مهمی را تشکیل می‌دهند. یکی از مسایل مورد توجه محققین پدیده خستگی و رشد ترک ناشی از آن است و با توجه به اینکه ترک و رشد آن سبب خرابی و شکست در ماشین آلات صنعتی می‌شود، بررسی رفتار ارتعاشی سازه‌های ترک‌دار از جمله تیرهای ترک‌دار ضروری می‌باشد.

تیر ترک‌دار از جمله عناصر مهم در ساختارهای دینامیکی است که توسط محققین مورد مطالعه قرار گرفته است. برای اولین بار Dimarogonas [1] روش تحلیلی را برای محاسبه پاسخ دینامیکی تیر ترک‌دار اویلر- برنولی پیشنهاد کرد، در این مدل ترک به عنوان فنر پیچشی در نظر گرفته شده است. Christides و Barr [2] تئوری پیوسته برای ارتعاش تیر اویلر- برنولی ترک‌دار را ارائه کردند. آنها از یک تابع اغتشاش نمایی استفاده نمودند و با بکارگیری اصل Hu-Washizu معادله حرکت تیر را استخراج کردند، سپس به بررسی رفتار ارتعاشی تیر ترک‌دار با تکیه‌گاه‌های ساده در دو انتها پرداختند. Shen و Pierre [3] مدل مشابهی برای بررسی ارتعاش تیر ترک‌دار ارائه کردند. آنها از روش المان محدود دو بعدی برای بدست آوردن تابع تمرکز تنش در محل ترک استفاده کردند. آن دو در پژوهش دیگری با استفاده از یک مدل پیوسته و بکارگیری اصل Hu-Washizu رفتار ارتعاشی تیر ترک‌دار را بررسی کردند[4]. معادله استفاده شده در مدل آنها به چند مقدار ثابت بستگی دارد که این مقادیر را با استفاده از نتایج روش المان محدود بدست آوردند. Carneiro و Inman [5] نشان دادند که معادله بدست آمده در مدل Shen و Pierre، معادله خودالحاق نمی‌باشد که این امر سبب بدست آمدن فرکانس‌های موهومی می‌شود که در شرایط عدم وجود استهلاک توجیه فیزیکی ندارد.

هدف اصلی از تحقیق حاضر، ارائه روش جدیدی برای بررسی رفتار دینامیکی تیر ترک‌دار است. در این مقاله مدل Shen و Pierre مورد بازنگری قرار گرفته و با استفاده از فرضیات واقع بینانه و روش کاملاً تحلیلی تابع اغتشاش ترک معرفی شده است. در بدست آوردن ثابت‌هایی که تابع اغتشاش به آنها بستگی دارد از روابط مکانیک شکست استفاده شده و در نتیجه معادلات از پیچیدگی کمتری برخوردارند و زمان محاسبات نیز کاهش یافته است.

پس از مدل کردن ترک و استخراج معادله حرکت توسط اصل همیلتون، به منظور بررسی اثر پارامترهای ترک (موقعیت و عمق ترک) بر رفتار ارتعاشی تیر ترک‌دار، معادله حرکت تیر با استفاده از روش گالرکین مورد تحلیل قرار گرفته است و نتایج بدست آمده، با نتایج تجربی در مرجع [6] مقایسه شده است.

## استخراج معادله حرکت تیر

تیر اویلر- برنولی دارای ترک لبه باز با تکیه‌گاه‌های ساده در دو انتها را مطابق شکل (۱) در نظر بگیرید:

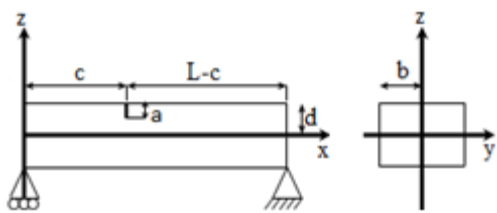

شکل(۱) : هندسه تیر اویلر- برنولی دارای ترک لبه باز

در شکل (۱)، $a$ مقدار عمق ترک، $d$ نصف عمق تیر، $L$ طول تیر، $b$ نصف عرض تیر و $c$ موقعیت ترک را نمایش می‌دهند.

طبق تئوری تیر اویلر برای تیر دارای ترک، کرنش طولی به صورت زیر بیان می‌شود[۴]:

$$\varepsilon_{xx} = (-z + \varphi(x,z))\frac{\partial^2 w}{\partial x^2} \quad (1)$$

در رابطه (۱) حرکت تیر در راستای محور $z$ با $w(x,t)$ نشان داده شده است و $\varphi(x,z)$ تابع اغتشاش ترک می‌باشد که در بخش بعدی معرفی خواهد شد.

انرژی کرنشی ذخیره شده در تیر از رابطه (۲) بدست می‌آید:

$$U = \iiint \frac{1}{2} E(\varepsilon_{xx})^2 dV \quad (2)$$

که در آن $E$، مدول الاستیسیته تیر می‌باشد. انرژی جنبشی تیر با در نظر گرفتن حرکت تیر صرفاً در راستای محور $z$ به صورت زیر می‌باشد:

$$T = \iiint \frac{1}{2} \rho (\frac{\partial w}{\partial t})^2 dx dy dz \quad (3)$$

با استفاده از اصل همیلتون معادله حرکت تیر به صورت زیر خواهد بود:

$$E\frac{\partial^2}{\partial x^2}(f(x)\frac{\partial^2 w}{\partial x^2}) + \rho A(\frac{\partial^2 w}{\partial t^2}) = 0 \quad (4)$$

در معادله حرکت بدست آمده تابع $f(x)$ به صورت زیر تعریف شده است:

$$f(x) = \iint (-z + \varphi(x,z))^2 dy dz \quad (5)$$

تابع تعریف شده در رابطه (۵) برای یک تیر سالم معادل با مقدار ممان اینرسی سطح یعنی $I$ خواهد بود، این مطلب با مقایسه معادله حرکت یک تیر سالم با رابطه (۴) نیز مشهود است.

با استفاده از نتایج تجربی می‌توان نشان داد که مقدار تابع $f(x)$ در محل ترک تقریباً برابر با ممان اینرسی سطح مقطع ناقص تیر در محل ترک است، به عبارتی:

$$f(c) = \int_{A_{crack}} z^2 dA \quad (6)$$

دو معادله دیگر بدست آمده از اصل همیلتون به صورت زیر می‌باشند:

$$\frac{\partial}{\partial x}(f(x)\frac{\partial^2 w}{\partial x^2})\delta w \Big|_0^L = 0 \quad (7\text{-}1)$$

$$f(x)\frac{\partial^2 w}{\partial x^2}\delta(\frac{\partial w}{\partial x})\Big|_0^L = 0 \quad (7\text{-}2)$$

از دو رابطه اخیر شرایط مرزی برای معادله حرکت (۴) به صورت زیر خواهند بود:

$$\begin{cases} x = 0 \rightarrow f(x)\dfrac{\partial^2 w}{\partial x^2} = 0, w = 0 \\ x = L \rightarrow f(x)\dfrac{\partial^2 w}{\partial x^2} = 0, w = 0 \end{cases} \quad (8)$$

## تابع اغتشاش ترک

توزیع تنش و کرنش توسط Tada [۷] و Paris و Sin [۸] مورد مطالعه قرار گرفته است. آنها به این نتیجه رسیده‌اند که توزیع تنش

یک تیر ترکدار به گونه‌ای است که در محل ترک مقدار تنش بیشینه می‌شود و با دور شدن از ترک مقدار تنش با نسبت ریشه دوم فاصله از ترک، کاهش می‌یابد. این اثر ترک و رفتار تنش توسط تابع اغتشاش به صورت زیر مدل شده است[۴]:

$$\varphi(x,z) = \left[z - m(z + \frac{a}{2})u(d - a - z)\right]e^{\frac{-\alpha}{d}|x - c|} \quad (9)$$

در این معادله پارامترهای $c, d, a$ در شکل (۱) نمایش داده شده‌اند و $u(d - a - z)$ تابع پله واحد است که به صورت زیر بیان می‌شود:

$$u(a - d - z) = \begin{cases} 1, & z < d - a \\ 0, & z \geq d - a \end{cases} \quad (10)$$

در معادله (۹) $\alpha$ و $m$ دو ضریب مقادیر مجهول هستند که در ادامه آنها را خواهیم یافت. $\alpha$ یک مقدار بی بعد ثابت و مثبت است که نرخ کاهش تنش با دور شدن از ترک می‌باشد. $m$ نیز ثابتی است که برابر با مقدار شیب خطی توزیع تنش در محل ترک است. وجود ترک باعث جابجایی اضافی در تیر می‌شود و در حالت کلی این جابجایی می‌تواند خیز یا شیب تیر باشد. در این مقاله تیر تحت خمش خالص $M$ قرار دارد و شیب اضافی در این حالت طبق قضیه کاستیلیانو به صورت زیر است[۹]:

$$\theta^* = \frac{\partial U_C}{\partial M} \quad (11)$$

که در آن $U_C$ انرژی کرنشی در محل ترک می‌باشد، این انرژی از معادله زیر بدست می‌آید[۱۰]:

$$U_C = \int_{crack} J(\gamma) d\gamma \quad (12)$$

که در آن $J(\gamma)$ از رابطه زیر بدست می‌آید[۱۰]:

$$J(\gamma) = \frac{1}{E^*}\left[(\sum_{i=1}^{6} K_{Ii})^2 + (\sum_{i=1}^{6} K_{IIi})^2\right] + \frac{1}{2G}(\sum_{i=1}^{6} K_{IIIi})^2 \quad (13)$$

در این معادله $G$ مدول برشی تیر است و در صورت استفاده از فرض تنش یکنواخت $E^* = E$ خواهد بود و با فرض کرنش یکنواخت به صورت $E^* = \dfrac{1}{1 - \upsilon^2}$ در نظر گرفته می‌شود، که در آن $\upsilon$ نسبت پواسون است. در این مقاله با کوچک در نظر گرفتن کرنش طولی از فرض کرنش یکنواخت استفاده کردیم. بعلاوه در معادله (۱۳)، $K_I, K_{II}, K_{III}$ ضرایب شدت تنش هستند [۱۱]. این ضرایب برای تیر ترکدار تحت خمش خالص به صورت زیر می‌باشد:

$$\begin{cases} K_I = \dfrac{Md}{I}\sqrt{\pi a}\,\beta_I(\gamma) \\ K_{II} = K_{III} = 0 \end{cases} \quad (14)$$

که در آن $\beta_I(\gamma)$ تابعی از عمق ترک است و به صورت زیر می‌باشد:

$$\beta_I(\gamma) = 1.12 - 1.4\gamma + 7.33\gamma^2 - 13.1\gamma^3 + 14\gamma^4 \quad (15)$$

و با خطای $\pm 0.2\%$ برای محدوده $0.6 > \gamma = \dfrac{a}{2d}$ صادق است.

۲

با قرار دادن معادله (۱۴) در معادله (۱۳) و استفاده از رابطـه (۱۱)، شیب اضافی ناشی از ترک به صورت زیر بدست می‌آید:

$$\theta^* = \frac{M(1-\upsilon^2)}{E}\phi(a,b,d) \quad (16)$$

که در آن:

$$\phi(a,b,d) = \frac{1}{bd^4}(153.9a^{10} - 324.1a^9 + 380.5a^8$$
$$- 272.4a^7 + 172.1a^6 - 88.1a^5 + 43.3a^4 - 11.1a^3 + 8.9a^2)$$

با مدل استفاده شده در قسمت قبل برای تیر ترک‌دار می‌توان نوشت:

$$\frac{d^2 w_c}{dx^2} = \frac{M}{Ef(x)} \quad (18)$$

با انتگرال‌گیری از رابطه (۱۸) خواهیم داشت:

$$\theta_c(x) = \frac{dw_c}{dx} = \frac{M}{E}\int \frac{1}{f(x)}dx + A \quad (19)$$

در این رابطه $A$ ثابتی است که با استفاده از شرایط مرزی تیر دو سر ساده بدست می‌آید و داریم:

$$\theta_c(x) = \frac{M}{E}\int \frac{1}{f(x)}dx - \frac{M}{EL}\int_0^L \int \frac{1}{f(x)}dxdx \quad (20)$$

با بدست آمدن $\theta_c(x)$ می‌توان گفت:

$$|\theta_c(c+\varepsilon) - \theta_c(c-\varepsilon)| = \theta^* \quad (21)$$

که در آن $\varepsilon$ مقدار کوچکی می‌باشد. با اسـتفاده از حـل عـددی دو معادله (۲۱) و (۶) می‌توان مقادیر $m$ و $\alpha$ را برای نسبت ترک با عمق‌های متفاوت بدست آورد. در این صورت تابع اغتشاش به صورت کامل تعریف می‌شود و می‌توان معادله حرکت (۴) را برای بدست آوردن فرکانس‌های طبیعی حل کرد.

**فرکانس‌های طبیعی**

معادله حرکت تیر طبق معادله (۴) بدست آمده است. این معادلـه دارای مقادیر ویژه حقیقی است ولی مسئله مقدار ویـژه بـه صـورت صریح برای آن قابل حل نیست چرا که ضریب $f(x)$ آن را غیرخطی ساخته است. بنابراین با استفاده از روش تقریبی گالرکین بـه دنبـال یافتن فرکانس‌های طبیعی هستیم. بـا اسـتفاده از روش جداسـازی متغیرها پاسخ معادله (۴) را به صورت زیر می‌نویسیم:

$$w(x,t) = X(x)T(t) \quad (22)$$

با جایگذاری رابطه (۲۲) در معادله حرکت تیر، داریم:

$$E\frac{d^2}{dx^2}(f(x)\frac{d^2 X}{dx^2}) = \lambda \rho A X(x) \quad (23)$$

در این شرایط $\sqrt{\lambda_n} = \omega_n$ فرکانس‌های طبیعی مـی‌باشـند. بـرای استفاده از گالرکین، $X(x)$ ترکیبی از توابع شکل مود را بـه صـورت زیر در نظر می‌گیریم:

$$X(x) = \sum_{n=0}^{N} a_n S_n(x) \quad (24)$$

تابع شکل مود مناسب برای تیر دو سر ساده به صورت زیر است:

$$S_n(x) = \sin(n\pi x / L) \quad (25)$$

تابع $T(t)$ برای پاسخ پایا به صورت زیر در نظر گرفته می‌شود:

$$T(t) = e^{-n\omega t} \quad (26)$$

با جایگذاری رابطـه (۲۲) در معادلـه (۴) و انتگرال‌گیـری از آن در طول تیر، مسئله مقدار ویژه به صورت زیر بدست می‌آید:

$$\det(K - \lambda_n M) = 0 \quad (27)$$

که در آن $K$ ماتریس سفتی با مرتبه $N \times N$ و درایه‌های زیر است:

$$k_{mn} = \int_0^L Ef(x)\frac{d^2 S_n(x)}{dx^2} \cdot \frac{d^2 S_m(x)}{dx^2}dx \quad (28)$$

$M$ نیز ماتریس قطری جرم با مرتبه $N \times N$ و درایه‌های زیر است:

$$m_{mn} = \int_0^L \rho A S_m(x)S_n(x)dx \quad (29)$$

در این مقاله مقدار $N$، برای حل معادله (۲۷)، ۱۰۰ در نظـر گرفتـه شده است.

**نتایج عددی**

مطابق واقعیت هر چه عمق ترک کم باشد، سرعت از بـین رفـتن اثـر آن بیشتر خواهد بود و عبارت $e^{-\frac{\alpha}{d}|x-c|}$ مقدار کمتری خواهد داشت. به عبارت دیگر با افزایش عمق ترک، مقدار $\alpha$ کم می‌شود، این رفتـار با نتایج بدست آمده در شکل (۲) هم‌خوانی دارد.

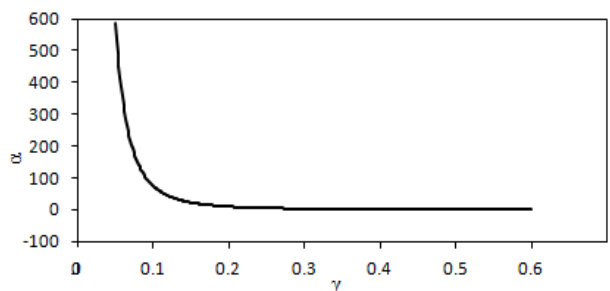

شکل(۲): تغییرات $\alpha$ (نرخ نمایی کاهش تنش) بر حسب عمق نسبی ترک

برای تأیید مدل ارائه شده، تیر آلومینیومی مورد بررسی قرار می‌گیرد و نتایج حاصل، با نتایج تست‌های تجربی موجود در پیشین، مقایسـه می‌شود. ابعاد هندسی و مشخصات مکانیکی تیر ترک‌دار آلومینیـومی دو انتها ساده، به صورت زیر است[۶]:
طـول تیـر $L = 235mm$ ، عمـق تیـر $2d = 7mm$، عـرض تیـر $2b = 23mm$، جـرم واحـد حجـم $\rho = 2800Kg/m^3$ و مدول یانگ $E = 72GPa$ است. همچنین پارامترهای بی‌بعد ترک عبارتند از: $\gamma = a/2d$ کـه نشـان دهنـده عمـق نسـبی تـرک و $\xi_c = c/L$ که نشان دهنده موقعیت نسبی ترک می‌باشد.
در شکل‌های (۳) تا (۵) منحنی تغییرات نسبت فرکانسی (نسبت فرکانس طبیعی تیر ترک‌دار به فرکانس طبیعی تیر سـالم) برحسـب عمق نسبی ترک در موقعیت‌های نسبی $\xi_c = 0.2$ و $\xi_c = 0.4$ رسم شده‌اند. این نمودارها نشان می‌دهد که فرکانس‌های طبیعی تیر ترک‌دار در موقعیت نسبی معین با افزایش عمق ترک کاهش می‌یابـد و اثر ترک با عمق معین، با نزدیک شدن موقعیت تـرک بـه گـره کـم می‌شود.

## نتیجه‌گیری

در تحقیق حاضر رفتار ارتعاش عرضی تیر ترک‌دار با مـدل پیوسـته تابع اغتشاش مورد بررسی قرار گرفت و با استفاده از مکانیک شکست تابع اغتشاش ترک بدست آمد. سپس بـا اسـتفاده از اصـل همیلتـون معادله حرکت تیر ترک‌دار استخراج گردید و فرکانس‌های طبیعی نیز با استفاده از روش گالرکین به دست آمد. نتایج نشان می‌دهد که بـه ازای موقعیت مشخص تـرک، بـا افـزایش عمـق تـرک فرکـانس‌هـای طبیعی کاهش می‌یابد. در نهایت صحت مدل ارائه شده با اسـتفاده از نتایج حاصل از تست‌های تجربی موجود به اثبات رسید.

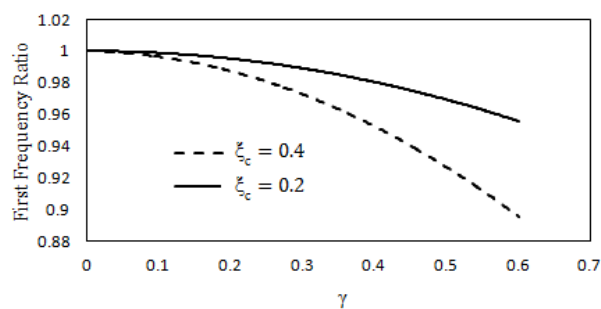

شکل(۳): نسبت کاهش فرکانس طبیعی اول بر حسب عمق نسبی ترک به ازای موقعیت نسبی ترک $\xi_c = 0.2$ و $\xi_c = 0.4$

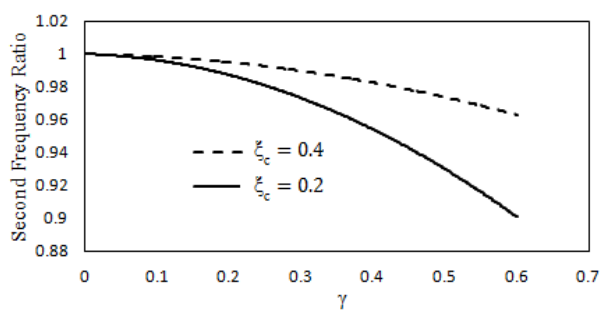

شکل(۴): نسبت کاهش فرکانس طبیعی دوم بر حسب عمق نسبی ترک به ازای موقعیت نسبی ترک $\xi_c = 0.2$ و $\xi_c = 0.4$

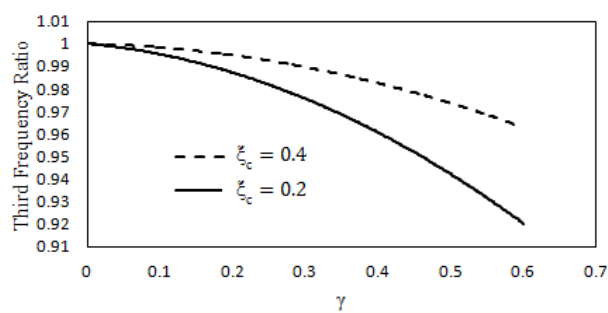

شکل(۵): نسبت کاهش فرکانس طبیعی سوم بر حسب عمق نسبی ترک به ازای موقعیت نسبی ترک $\xi_c = 0.2$ و $\xi_c = 0.4$

در شکل (۶) منحنی کاهش فرکانس طبیعی اول در مقابل عمق ترک به ازای $\xi_c = 0.5$ رسم شده و با نتایج تجربی مقایسه شـده اسـت. در این مقایسه حداکثر خطا ۲/۷٪ می‌باشد. این نتایج نشان می‌دهـد که مدل ارائه شده از دقت خوبی برخوردار است.

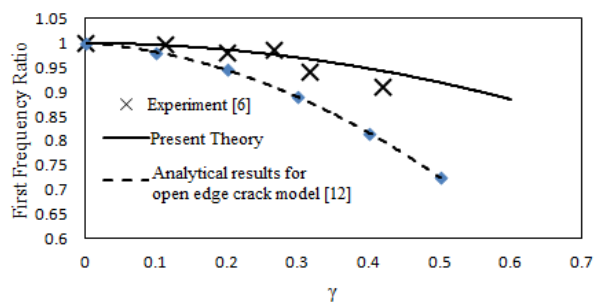

شکل(۶): نسبت کاهش فرکانس طبیعی اول برحسب عمق نسبی ترک در مقایسه با نتایج تجربی [۶] و مدل ارائه شده در مرجع [۱۲]، به ازای موقعیت نسبی ترک $\xi_c = 0.5$.